Autocorrelations of random fractal apertures and phase screens

Jonathan F. Schonfeld
Lexington, Massachusetts 02420
jschonfeld@aya.yale.edu

Running head: Random fractal apertures

Abstract

We introduce a new product representation for general random binary fractal apertures defined by removing voids from Euclidean space, and use it to derive a simple closed-form expression for ensemble-averaged correlations. Power-law scaling at short distance follows almost immediately. Similar techniques provide easy constructions of objects with fractional Brownian short-distance behavior for phase screens and other applications.

## 1. INTRODUCTION

Fractal optics has been an active subject of research for nearly forty years [1,2]. A central theme is that propagation from fractal apertures or through fractal media induces characteristic behavior in optical fields. This is generally established by explicitly assuming that fractal apertures or media themselves exhibit power-law correlations, or experimentally creating fractal apertures and measuring resulting fields [3], or carrying out laborious analyses for very specific fractal apertures [4].

This paper introduces a new product representation of random binary fractals defined by removing voids from Euclidean space that makes it easy to calculate ensemble-averaged aperture autocorrelation functions in closed form, from which short-distance power-law scaling follows almost immediately. Similar techniques permit easily constructing objects with short-distance fractional Brownian behavior, which are valuable for emulating turbulent propagation media in the laboratory with phase screens and also for describing traffic on communications networks [7].

## 2. PRODUCT REPRESENTATION AND ENSEMBLE-AVERAGED AUTOCORRELATION

For the purposes of this paper, a random "take-away" fractal is a set formed by the following recursive procedure: Start with a Euclidean space of arbitrary dimension $D$, and an arbitrary reference void of volume (area for $D=2$) $V$. Distribute points randomly throughout the Euclidean space with arbitrary density $\rho$, and, centered at each such point, remove a copy of the reference void. Call this the zeroth iteration. Now choose an arbitrary scale factor $\xi>1$ and define the $k$'th iteration inductively as follows:

- Distribute points randomly with density $\rho\xi^{Dk}$ throughout whatever part of the Euclidean space has not been removed by preceding iterations.
- Centered at each such point, remove from the $k$-1'st iteration a copy of the reference void linearly scaled by factor $\xi^{-k}$.

In the limit of infinite $k$, what's left has fractal dimension $D+\ln(1-\rho V)/\ln\xi$ [4]. The factor $(1-\rho V)$ is the volumetric proportion of iteration $k$-1 removed by iteration $k$. For small $\rho V$ the fractal dimension can be written as $D-\varepsilon$ where $\varepsilon = \rho V/\ln\xi$. (The geometry of the reference void can have its own probability distribution, but this is more generality than we require.)

For small $\rho V$ it is unimportant that voids removed at iteration $k$ might overlap with one another or with voids removed at earlier iterations. In fact, the possibility of overlap allows us to replace the phrase, "whatever part of the Euclidean space has not been removed by preceding iterations" by "the Euclidean space," without any meaningful change to the outcome of the fractal construction. This is the central insight of this paper.

The binary aperture function $\Phi$ is defined as zero in the voids and one elsewhere (we ignore overall aperture diameter). The ensemble-averaged autocorrelation is $<\Phi(\boldsymbol{x})\Phi(\boldsymbol{x}+\boldsymbol{y})>$ (appropriately normalized) for arbitrary $\boldsymbol{x}$ and $\boldsymbol{y}$. Its Fourier transform is the far-field intensity pattern. To exploit the central insight, define the function $H$ to be one inside the reference void and zero elsewhere. Then after the zeroth iteration, $\Phi(x)$ is just a product of factors $1-H(x-z)$, one for each point $z$ in the iteration's random distribution. In the same way, iteration $k$ simply multiplies $\Phi$ by another such product, this time of factors $1-H(\xi^k(x-z))$, one for each point $z$ in iteration $k$'s random distribution. Since all $z$'s in all iterations are independent random variables, the autocorrelation factors in a similar way. After the zeroth iteration, the autocorrelation is a product of factors $<(1-H(x-z))(1-H(x+y-z))>$, one for each $z$ in the iteration's random distribution. Iteration $k$ multiplies this result by a product of factors $<(1-H(\xi^k(x-z)))(1-H(\xi^k(x+y-z)))>$, one for each $z$ in iteration $k$'s random distribution.

Averaging each autocorrelation factor amounts to integrating with respect to $z$ over all Euclidean space and dividing by the total volume of Euclidean space. Since $H$ has compact support but total volume is infinite, each autocorrelation factor on its own devolves to unity. But the number of factors is also proportional to total volume, so the product definition of the exponential implies that iteration $k$ in total contributes the following factor to the overall autocorrelation:

$$\exp\left[-2\rho\xi^{kD}\int H(\xi^k x)d^D x + \rho\xi^{kD}\int H(\xi^k x)H(\xi^k(x+y))d^D x\right] = \exp[-2\rho V + \rho C(\xi^k y)] \quad (1)$$

where

$$C(y) = \int H(x)H(x+y)d^D x. \quad (2)$$

So, up to a vanishing denominator normalization – factor of $\exp(-2\rho V)$ for each iteration – the complete autocorrelation in closed form is

$$<\Phi(x)\Phi(x+y)> = \exp[\rho \sum_{k=0}^{\infty} C(\xi^k y)]. \quad (3)$$

The vanishing normalization has a simple interpretation: The product of $\exp(-\rho V)$, one for each iteration, is just the fraction of Euclidean space left after the fractal construction. The product of $\exp(-2\rho V) = [\exp(-\rho V)]^2$ is just the fractional overlap between two copies of the fractal shifted far enough apart to be uncorrelated.

To derive behavior for small $y$, write $y=ru$, where $u$ is a unit vector. Then the finite extent of $H$ implies $C=V$ for $r=0$ and $C=0$ for $r$ greater than some $u$-dependent value $r_u$ (for a spherical void, $r_u$ is $u$-independent). So for $r > r_u$, the sum in Eq. (3) is zero, while for small r, the sum in Eq. (3) is roughly (up to a remainder that becomes fractionally insignificant as $r$ approaches zero) $C(0)=V$ times $k_{max}$, the number of values of $k$ for which $\xi^k r < r_u$, as long as $C$ interpolates gradually between $r=0$ and $r_u$. Thus for small $y$ the exponent in Eq. (3) is roughly $\rho V \log_\xi(r_u/r)$, leading to the power law

$$(r_u/r)^{(\rho V/\ln \xi)} = (r_u/r)^\varepsilon. \quad (4)$$

To understand how gradually $C$ actually interpolates, note that in general, for small $r$,

$$C(y) \sim C(0) - \frac{1}{2}r\int_{surface}|\hat{s}\cdot\hat{u}|ds \quad (5)$$

where the integral covers the entire surface (circumference for $D=2$) of the reference void, $ds$ is surface element and $\hat{s}$ is unit surface normal. Since $C(0) = V$, Eq. (5) says fractional slope at $r = 0$ is roughly volume-to-surface area, which for a simple convex void is O($r_u$), as gradual as can be. The situation can be very different for more complex shapes, for example hyperbolic spheres (the invariant shapes of special relativity), appropriately truncated, for which surface/volume is much more exaggerated. In that case $C$ hardly varies at all below $r_u$ until $r$ gets extremely close to 0.

### 3. FRACTIONAL BROWNIAN BEHAVIOR AND PHASE SCREENS

It is easy to change the rules so Eq. (4) is replaced by small-distance scaling with a positive power of $y$. Simply assume that at every iteration $k$, density evolves as $\rho \xi^{(D-\beta)k}$ for some positive $\beta$. Then for small $y$ (large $k_{max}$), the right-hand-side of Eq. (3) becomes

$$\sim \exp\left[\rho C(0) \sum_{k=0}^{k_{max}} \xi^{-\beta k}\right]$$
$$= \exp\left[\frac{\rho V(1-\xi^{-\beta(k_{max}+1)})}{1-\xi^{-\beta}}\right]$$
$$\sim 1 - \frac{\rho V \xi^{-\beta k_{max}}}{\xi^\beta - 1}$$
$$\sim 1 - \frac{\rho V}{\xi^\beta - 1}(r/r_u)^\beta \quad (6)$$

where the last two expressions ignore overall multiplier $\exp[\rho V \xi^\beta/(\xi^\beta - 1)]$. The coefficient of $(r/r_u)^\beta$ in Eq. (6) also ignores a jittering correction that accounts for the fact that $C(\xi^k y)$ may not be approximately equal to $C(0)$ for the last few terms in the sum in Eq. (6); analyzing this jitter is beyond the scope of this paper. Since all the scaled correlations $C$ in Eq. (3) vanish for $r > r_u$, Eq. (6) implies, again up to normalization,

$$< [\Phi(x) - \Phi(x+y)]^2 > \sim 2(r/r_u)^\beta \frac{\rho V}{\xi^\beta - 1} \exp\left[\frac{\rho V \xi^\beta}{\xi^\beta - 1}\right] \quad (7)$$

for small $y$, i.e. fractional Brownian behavior. Alternatively, one could let density scale as in Sec. 2, and simply multiply the void function $H$ for iteration $k$ by $e\xi^{\beta k}$ for some parameter $e$; the result is the same as in Eq. (7), replacing $\rho$ by $\rho e^2$. In this case, $\Phi$ is no longer binary, and introducing a factor 1-$eH$ into $\Phi$ doesn't carve a localized void from an aperture (or reset some ones to zeros in a binary set), but rather imprints a localized aberration on a phase screen. Multiplicative imprinting is difficult to implement in real hardware; fortunately, similar logic shows that Eq. (7) without the exponential multiplier also characterizes the additive (printable) process $\Phi = 1-\Sigma eH$, where the sum has one contribution $eH(\xi^k(x-z))$ for every iteration $k$ and every point $z$ in iteration $k$'s random distribution. A spray process for creating Kolmogorov-like phase screens is described in [6]. It would be interesting to understand the droplet distribution in [6] and see if it can be related to the concepts in this paper.

1. M. V. Berry, J. Phys. A: Math. Gen. **12**, 781–797 (1979).
2. M. Segev, M. Soljačić and J. M. Dudley, Nature Photonics **6**, 209–210 (2012).
3. Y. Kim and J. L. Daggard, Proc. IEEE **74**, 1278 (1986).